\documentclass[aps,prl,showpacs,twocolumn,superscriptaddress]{revtex4}
\usepackage{amsmath}
\usepackage{amsfonts,dsfont}
\usepackage{amssymb}
\usepackage{graphicx}
\usepackage{natbib}
\usepackage[colorlinks=true,linkcolor=blue,urlcolor=black,citecolor=blue]{hyperref}

\newcommand{\charge}{Q}

\usepackage{mathptmx, courier, pifont}
\usepackage[scaled=0.92]{helvet}
\usepackage[T1]{fontenc}
\usepackage{textcomp}

\begin{document}

\title{Quantum simulator for many-body electron--electron Coulomb interaction 
with ion traps}

\author{Da-Wei Luo}
\affiliation{Beijing Computational Science Research Center, Beijing 100084, China}
\affiliation{Department of Theoretical Physics and History of Science, 
The Basque Country University UPV/EHU, 48080 Bilbao, Spain}
\affiliation{Ikerbasque, Basque Foundation for Science, 48011 Bilbao, Spain}

\author{P. V. Pyshkin}
\affiliation{Beijing Computational Science Research Center, Beijing 100084, China}
\affiliation{Department of Theoretical Physics and History of Science, 
The Basque Country University UPV/EHU, 48080 Bilbao, Spain}
\affiliation{Ikerbasque, Basque Foundation for Science, 48011 Bilbao, Spain}

\author{Michele Modugno}
\affiliation{Department of Theoretical Physics and History of Science, 
The Basque Country University UPV/EHU, 48080 Bilbao, Spain}
\affiliation{Ikerbasque, Basque Foundation for Science, 48011 Bilbao, Spain}

\author{Mike Guidry}
\affiliation{Department of Physics and Astronomy, University of Tennessee, 
Knoxville, Tennessee 37996, USA
}


\author{J. Q. You}
\affiliation{Beijing Computational Science Research Center, Beijing 100084, China}

\author{Lian-Ao Wu}
\email{lianaowu@gmail.com}
\affiliation{Department of Theoretical Physics and History of Science, 
The Basque Country University UPV/EHU, 48080 Bilbao, Spain}
\affiliation{Ikerbasque, Basque Foundation for Science, 48011 Bilbao, Spain}

\date{\today}

\begin{abstract}
We propose an analog quantum simulator that uses ion traps to realize the 
many-body electron--electron Coulomb interaction of an electron gas. This 
proposal maps a system that is difficult to solve and control to an 
experimentally-feasible setup that can be realized with current technologies. 
Using a dilatation transform, we show that ion traps can efficiently simulate 
electronic Coulomb interactions. No complexity overhead is added if only the 
energy spectrum is desired, and only a simple unitary transform is needed on the 
initial state otherwise. The runtime of the simulation is found to be much 
shorter than the timescale of the corresponding electronic system, minimizing 
susceptibility of the proposed quantum 
simulator to external noise and decoherence. 
This proposal works in any number of dimensions, and could be 
used to simulate different topological phases of electrons in graphene-like 
structures, by using ions trapped in honeycomb lattices.
\end{abstract}


\maketitle

{\it Introduction.}{\bf--}
Many-body interactions are fundamental to understanding a variety of interesting 
phenomena.  Quantum many-body problems are notoriously difficult to solve for 
the full energy spectrum, and are even more problematic for dynamical 
properties.  Few realistic exactly-solvable models exist, necessitating  
approximations that typically are valid only in some regions of the parameter 
space. The root of this difficulty is that the Hilbert space grows 
exponentially as the number of quantum particles increases~\cite{Feynman1982}, 
so on a classical computer the resources and time required to solve a problem 
exhibit a corresponding exponential growth. Some powerful numerical tools such 
as quantum Monte Carlo methods and the density matrix renormalization group have 
had some success for some systems, but the complexity involved prevents an 
efficient numerical study of many other interesting problems, especially for 
quantum dynamics and higher-dimension systems.

A powerful alternative to calculation is  \emph{quantum 
simulation}~\cite{Feynman1982, Georgescu2014,Cirac2012}. A universal quantum 
computer has not yet been realized but quantum simulations of specific systems 
have enjoyed considerable success, aided by experimental advances such as the 
realization of high-fidelity quantum gates and increasingly precise 
measurement~\cite{Harty2014,Garcia-Ripoll2003}. They have been used to study 
quantum phase transitions~\cite{Retzker2008}, open quantum 
systems~\cite{Piilo2006}, and pairing Hamiltonians~\cite{Wu2002}, to name just a 
few applications. Quantum simulators follow the laws of quantum mechanics, with 
an exponentially-growing computing capability~\cite{Feynman1982,Georgescu2014} 
that can match the exponential growth of problem size with particle number. The 
idea to simulate one quantum system with another has been 
proven~\cite{Lloyd1996} to be efficient, at least for any many-body system 
having only few-body particle correlations. 

A quantum simulator works by designing a custom Hamiltonian $H_\mathrm{s}$ so 
that the evolution operator $U=\widehat{T}\exp[-i\int_0^t H_\mathrm{s}d\tau]$ 
behaves like the physical system one wishes to simulate. Then a precise 
measurement at the end of the time evolution gives the physical quantity of 
interest. A useful quantum simulator requires  high-fidelity 
Hamiltonian engineering and initialization, and precise measurement. 
A number of systems have been realized experimentally to implement the quantum 
simulation task, including ion traps, ultracold atoms in optical lattices, NMR 
nuclear spins, and superconducting qubits.

Many-body electron--electron (e--e) Coulomb interactions play a critical role in 
many important phenomena such as the fractional quantum Hall 
effect~\cite{Wu2014,Stormer1999} and high $T_c$ 
superconductors~\cite{Rietschel1983}. Dealing with the mutual Coulomb 
interactions between all the electrons is daunting but essential to a deep 
understanding of such systems. In this Letter, we propose to use quantum 
simulations to obtain the properties of systems exhibiting many-body e--e 
Coulomb interactions. The 
quantum simulator may be realized using an ion 
trap~\cite{Blatt2012,Benassi2011,Georgescu2014}.  Control methods for ion traps 
may be implemented with current technologies~\cite{Garcia-Ripoll2003,Harty2014} 
and they have been used widely in quantum simulation and quantum information 
tasks~\cite{Haffner2008155}, with applications in areas such as quantum 
chemistry~\cite{Yung2014} and mass spectrometry~\cite{Douglas2005}.  

Electrons in the physical system to be emulated and ions in the trap carry 
different masses and charges, so a direct simulation is not possible. However, 
we have employed a dilatation transform to establish an explicit mapping  
between the simulator on one timescale and the interacting electrons on a 
different timescale. Therefore, the ion trap at rescaled times can be used to 
simulate the e--e Coulomb interactions at physical times. To read out the 
results, we propose imaging of the ions to record their positions, which then 
have a direct mapping to the positions of the interacting electrons. Remarkably, 
if only the energy spectrum is desired the unitarity of the dilatation transform 
implies that it will not alter the spectrum so it is not necessary to generate 
the transform physically. 

This proposal is dimensionality-agnostic. For example, 
it could be employed for simulating the behavior of 2D electron gases.
For charged particles confined in 2D, some interesting phenomenon can arise, 
such as various quantum Hall effects~\cite{Wu2014,Stormer1999}.
Moreover, by loading the ions in honeycomb lattices, one could also 
realize different topological phases characterizing the behavior of electron in 
graphene-like structures \cite{Polini2013}. The ion trap simulator allows 
for a study of the finite-size boundary effects for the interacting electrons, 
where bound states are possible.
As a concrete example of this proposal, we shall 
illustrate the experimental setup of the quantum simulation by using calcium 
ions carrying one positive charge.

{\it The interacting electron gas.}{\bf--}
An interacting electron gas can be described by the Hamiltonian 
($\hbar\equiv 1$)~\cite{Bohm1953}
\begin{equation}
	H_{\rm eg}=\sum_{i=1}^N \frac{{\pmb{p}}_i^2}{2m_{\rm e}}
	+\sum_{i <j}^N \frac{e^2}{|{\pmb{q}}_i-{\pmb{q}}_j|}, \label{eq_he}
\end{equation}
where $m_{\rm e}$ is the electron mass and ${\pmb{p}}_i$ and ${\pmb{q}}_i$ are 
the momentum and position operators, respectively, for the $i$-th electron. The 
first term represents the kinetic energy and the second term represents the 
potential energy resulting from mutual Coulomb interactions among the electrons. 
 Because of the many-body nature of the problem, the Hilbert space grows 
exponentially with electron number, making an exact solution of the 
Schr\"odinger equation difficult. Although in some cases it may be possible to 
consider only the average effect of the electrons (mean field approximations), a 
complete description requires accounting for the interaction between all 
electrons.

Traditionally, perturbation theory approximations or numerical tools have been 
used to tackle this problem, with mixed success. In this Letter, we take a 
fundamentally different approach:  we propose an experimentally-feasible quantum 
simulator that can faithfully represent the Coulomb interaction between the 
electrons, thus bypassing cumbersome calculations.

{\it Ion-trap simulator.}{\bf--}
Typically quantum simulation consists of three stages: (1)~preparation of an 
initial state, (2)~time evolution under a specifically-engineered Hamiltonian, 
and (3)~readout of the result, which can be achieved through quantum phase 
estimation procedures or measurement. Assuming the quantum simulator to be 
well-controlled experimentally, the most critical task is to design a 
Hamiltonian of the quantum simulator leading to the required propagator, so that 
a map exists between the initial and final states of the simulator 
and those of the system under consideration~\cite{Georgescu2014}. Ion traps are
widely-used and experimentally well-controlled systems, with the ions either 
localized or undergoing axial and cyclotron motions as in Penning 
traps~\cite{Brown1986}. Since the fermionic ions exhibit mutual Coulomb 
interaction, an 
ion trap loaded with identical ions may be a good candidate for the simulation 
of e--e Coulomb interactions.

\begin{figure}[t]
	\centering
	\includegraphics[width=0.7\columnwidth]{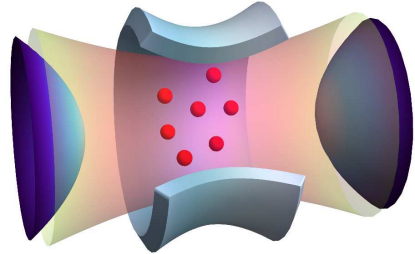}
	\caption{(Color online) Schematic representation of the ion-trap 
simulator for Coulomb interactions: $N$ identical ions, mutually interacting via 
the Coulomb force.
}
	\label{fig_model}
\end{figure}

To simulate the electron gas, we propose an analogue quantum 
simulator using an ion trap with $N$ identical ions interacting by Coulomb 
interactions, as illustrated schematically in Fig.~\ref{fig_model}. 
The Hamiltonian is given by
\begin{equation}
	H_{\rm s}=\sum_{i=1}^N \frac{{\pmb{p}}_i^2}{2m_{\rm ion}}
	+\sum_{i <j}^N \frac{\charge ^2 e^2}{|{\pmb{q}}_i-{\pmb{q}}_j|},
	\label{eq_hsim}
\end{equation}
where $m_{\rm ion}$ is the mass of the ion and $\charge $ is the degree of 
ionization. 
This Hamiltonian is formally similar to Eq.~\eqref{eq_he} but there are two 
significant differences: (1)~The ion to electron mass ratio is of order 
$10^4-10^5$ for ions such as $^9 \text{Be}^+$ or $^{111} 
\text{Cd}^+$ commonly used in traps, and $\charge $ may be greater than 
$1$. Since the kinetic term and the Coulomb interaction terms do not 
commute, these disparities rule out direct use of the ion trap as an 
analog simulator for this problem. For the ion trap to faithfully simulate the 
e--e Coulomb interaction, an explicit mapping between the two systems is 
required. We shall now show that this can be achieved by the introduction of a 
scaled evolutionary time for the simulator.

\textit{Dilatation operator.}
Let us consider the (unitary) dilatation operator~\cite{Wybourne1974}
\begin{equation}
S(\pmb{r})=\exp \left[\sum_{j=1}^N \frac{ir_j}{2} \left( 
{\pmb{q}}_j\cdot{\pmb{p}}_j+{\pmb{p}}_j\cdot{\pmb{q}}_j\right) 
\right],\label{eq_dil_trans}
\end{equation}
where $\pmb{r}=(r_1,\ldots,r_N)$, with $r_k$ being a 
real dilatation parameter for the $k$th particle.  Using the 
Baker--Campbell--Hausdorff formula 
\begin{equation}
	\exp[\alpha A] B \exp[-\alpha A]=B+\sum_{m=1}^{\infty}
	\frac{\alpha^m}{m!} 
[_m A,B]
\end{equation}
where $[_m A,B]=[A,[_{m-1} A,B]]$ and $[_1 A,B]=[A,B]$ is the commutator, 
one finds that for the $k$th particle the position and momentum operators 
transform as
\begin{equation}
S ^\dagger (\pmb{r}) {\pmb{q}}_k S(\pmb{r})=\exp({-r_k}){\pmb{q}}_k, 
\qquad
S ^\dagger (\pmb{r}) {\pmb{p}}_k 
S(\pmb{r})=\exp({r_k}){\pmb{p}}_k
\label{eq_dil_bch} .
\end{equation}
Thus the dilatation 
transform scales the momentum and position terms differently, allowing the 
ratio between the two terms in the Hamiltonian \eqref{eq_hsim} to be tuned. 

We take $r_k\equiv r$ for $k=1,\ldots,N$ and denote the corresponding 
dilatation transform as $S(r)$. Because  $S(r)$ is unitary, 
$$
S ^\dagger 
(r) f_k({\pmb{p}}_k,{\pmb{q}}_k) S(r)=f_k(\exp(r) {\pmb{p}}_k, \,\exp(-r) 
{\pmb{q}}_k)
$$
for any operator functions $f_k$ and
\begin{equation}
	\widetilde{H}_{\rm s} \equiv S ^\dagger (r) H_{\rm s} S(r)
	=\exp(r) \charge ^2 \left[ \sum_{i=1}^N \frac{{\pmb{p}}_i^2}{2m_{\rm 
eff}}
	+\sum_{i <j}^N 
\frac{e^2}{|{\pmb{q}}_i-{\pmb{q}}_j|}\right],
\label{eq_hsimT}
\end{equation}
where $m_{\rm eff}=\exp(-r) \charge ^2 m_{\rm ion}$ represents the the effective 
mass 
after the dilatation. 
Then, by requiring $m_{\rm eff}=m_{\rm e}$, one recovers exactly the
e--e Hamiltonian in Eq.~\eqref{eq_he}. 
This relation fixes the dilatation parameter $r$
and the scaled runtime $\tilde{t}=t/(\exp(r) \charge ^2)$, so that the 
evolution operator is
\begin{equation}
U(t)=\exp({-iH_{\rm eg}t})=S ^\dagger (r) 
\mathcal{U}\left(\tilde{t}\right) S (r),
\;
\mathcal{U}\left(\tilde{t}\right)=\exp({-i {H}_{\rm s}\tilde{t}} ).
\end{equation} 
This establishes a one-to-one mapping between the initial state 
$|\psi(0) \rangle$ of the interacting electrons and the initial state  $|\phi(0) 
\rangle$ of the ion-trap quantum simulator, as well as between the corresponding 
final states $|\psi(t_{\rm f}) \rangle$ and $|\phi(\tilde{t}_{\rm f}) \rangle$, 
so that quantum simulation of the e--e Coulomb interactions using ion traps is 
possible. 

Physically the map is a rotation described by the dilatation operator and a 
corresponding rescaling of time for the propagator. The dilatation parameter 
and the new time scale are determined solely by the mass ratio between the ion 
and electron, and the degree of ionization. As a bonus, since the mass of the 
electron is much less than that of the ions and $\charge  \geq 1$, we have 
$\charge ^2 \exp(r) 
\gg 1$. Therefore, to simulate the propagator $U(t)$ the simulator runtime 
$\tilde{t}$ is much less than the physical runtime $t$. This efficiency is 
particularly beneficial since shorter runtimes decrease the susceptibility of 
the quantum simulator to external noise and decoherence. When we are 
interested in boundary effects or need to take into consideration the finite 
size nature of the simulator, the dilatation parameter also dictates the mapping 
between the boundaries of the electron gas and the ion trap size. For example, 
with a hard wall boundary of width $w$ for the electron gas, the dilation 
transform will map that to a hard wall boundary of width $\exp(-r) w$ for the 
ion trap.
A similar scaling would apply to any external potential (as in the case 
of an honeycomb lattice), namely $V_s(\pmb{q})=\charge ^2 \exp(r) 
V_\mathrm{ext}(\exp(r) 
\pmb{q})$, where $V_s$ and $V_\mathrm{ext}$ are the potentials for the ion-trap 
simulator and electron gas, respectively.

\textit{Readout.} 
Results may be read out using an imaging technique to measure the 
position of the ions~\cite{Preiss2015} in the trap and build up a history of 
the position over time. Assume that one prepares an  arbitrary initial state of 
the simulator as $|\varphi(0) \rangle=\int c_v |v \rangle$, where $|v \rangle$ 
labels the eigenvector with an eigenvalue of $E_v$. Then, a position measurement 
would return 
\begin{align*}
\langle n(\pmb{q}) \rangle&=\int dv dv' c_v^* c_{v'} 
\exp[{i(E_v-E_{v'})t}] \langle v|n(\pmb{q})|v' \rangle
\\
&\equiv\int dv dv' F(v,v') 
\exp[{i(E_v-E_{v'})t}],
\end{align*}
where $n$ is the density of state at position $\pmb{q}$. The dilatation 
transformation is unitary, so 
that $U(t)$ and $\mathcal{U}\left(\tilde{t}\right)$ have the {\em same 
spectrum,} which can be extracted by means of a 
simple Fourier transformation~\cite{Wu2002}, 
$$
\bar{S}(\omega)=\int dv dv' 
\widetilde{F}(v,v')\delta[\omega-(E_v-E_{v'})]
$$
with the sharpness of the delta 
function being related to the sampling frequency of the measurements.

In principle, if one would like to simulate the dynamical evolution 
of a precise electronic initial state $|\psi(0)\rangle$, one should prepare the 
corresponding initial state of the simulator as $|\varphi(0) 
\rangle=S(r)|\psi(0)\rangle$. This could be achieved by propagating the initial 
state $|\psi(0) \rangle$ using the Hamiltonian 
\begin{equation}
H' = -\tfrac12 \sum_{j=1}^N \left( 
{\pmb{q}}_j\cdot{\pmb{p}}_j+{\pmb{p}}_j\cdot{\pmb{q}}_j\right)\label{h_s_rot}
\end{equation}
for a time 
duration of $r$, according to Eq.~\eqref{eq_dil_trans}. We then let the state
propagate for a duration of $t$, followed by an inverse propagation of~\eqref{h_s_rot}.
Then, a
measurement of the ions would give expectation values for the electron gas 
according to
$$
\langle 
\varphi(0)|\mathcal{U}^\dagger(\widetilde{t}) O  
\mathcal{U}(\widetilde{t})|\varphi(0)\rangle
=\langle \psi(t)| O |\psi(t) \rangle,
$$
where $O$ is the observable under consideration and $\mathcal{U}
(\widetilde{t})=S(r) U(t) S ^\dagger(r)$.

We also envisage a more general way to obtain the spectrum, 
through the quantum phase estimation algorithm~\cite{Abrams1999,Nielsen2000}. 
This would require the coupling of the simulator to a quantum circuit capable of 
generating a controlled-U operation that takes a qubit state as control and 
applies the unitary operation on the wave function only if the control qubit is 
in the $| 1 \rangle$ state. For an estimated phase of $n$-bit precision, one 
needs $n$ Hadamard gates to transform the $n$ ancillary qubits from $|0 \rangle$ 
to $|+ \rangle=(|0 \rangle+|1 \rangle)/\sqrt{2}$.  Denoting the controlled-U 
operation as 
$$
K_j=|1 \rangle\langle 1|_j  \mathcal{U}^{2^{j-1}}(\tilde{t})+|0 
\rangle\langle 0|_j \mathds{1} ,
$$ 
for the $j$th qubit
$|{{q}}_j \rangle=c_0 
|0 \rangle_j+c_1 |1 \rangle_j$,
we have
$$
K_j  |{{q}}_j \rangle |\psi 
\rangle=[c_0 |0 \rangle_j+c_1 \exp({i2\pi\varphi {2^{j-1}}})  |1 \rangle_j]\otimes 
|\psi \rangle
$$
where 
$\mathcal{U}(\tilde{t})|\psi \rangle=\exp({i2\pi\varphi}) 
]|\psi \rangle$. 
The phase $\varphi$ is what one should read out for the 
simulation. Treating the product basis for the qubits $|v_1,v_2,\ldots,v_n 
\rangle$, where $v_i=\{0,1\}$ as an $n$ bit binary number $b$ so that the basis 
can be denoted as $|b \rangle$, $b=0,\ldots, 2^{n-1}$, the inverse quantum 
Fourier transform (IQFT) performs the mapping 
$$
\sum_{j=0}^{2^n-1} \frac{\exp({i2\pi 
j\varphi})}{2^{n/2}}  |j \rangle|\psi \rangle \rightarrow |M \rangle|\psi 
\rangle ,
$$
where $M$ denotes an $n$-bit estimate for the phase $\varphi$ after 
measurement.

{\it Experimental setup.}{\bf--}
A realistic experimental setup for simulation of e--e Coulomb interactions is 
afforded by trapped $^{40}{\rm Ca}^+$  ions~\cite{Blatt2012}. The number of 
$^{40}{\rm Ca}^+$ ions that can be loaded with present technology ranges from a 
few to tens of thousands~\cite{Herskind2009}, and individual addressing of the 
ions has  been achieved~\cite{Blatt2012}. From the mass ratio between a calcium 
ion and an electron, the timescale  $\tilde t$ for this simulator is related to 
the physical electronic timescale $t$ by $\tilde{t}\approx 1.37\times 
10^{-5}t$. 

To simulate an $N$-electron system the trap is loaded with $N$ 
ions. The duration for 
propagation  depends on the precision required in the simulation but is limited 
by the trap decay time, which is around $1 \mu$s for  a radio-frequency ion trap 
that incorporates an optical cavity~\cite{Herskind2009}.   Since the physical 
electronic time in this example is about $10^5$ times the trap evolution time, a 
relatively long electronic time ($\sim 0.1$ s) can be attained for 
a trap evolution time that is less than the trap decay time.

To read out the simulation results the ions can be imaged to build a record of 
position measurements.  Then the records can either be mapped to the real 
electron positions for appropriate initial states, or a Fourier transform can be 
used to obtain the energy spectrum. Alternatively, a phase-estimation 
algorithm with a phase precision of $n$ bits can be implemented by sending $n$ 
qubits initially prepared in the $|0 \rangle$ state through Hadamard gates, 
which is also realizable using ion traps~\cite{Harty2014,Kuvshinov2005}. This 
readout procedure is capable of high precision since the realization IQFT can 
be scalable in a semi-classical way~\cite{Griffiths1996,Zhou2013,Ad}.

{\it Conclusion.}{\bf--}
We have proposed a dimensionality-agnostic quantum simulation of the Coulomb interactions 
in an electron gas by using ion traps loaded with positive ions. The disparity 
between the masses of electrons and ions, as well as the different charges that 
the ions may carry, preclude a direct simulation. However, we have shown through 
a dilatation transform that the propagator of the electron gas at time $t$ gives 
a spectrum that is mapped one-to-one to the spectrum of the ion trap at a 
rescaled time $\tilde t$, with the scaling factor between $\tilde t$ and $t$ 
specified completely by the mass of the ion and its degree of ionization. An 
imaging on the ions can be used to build a measurement record of their 
positions, which is mapped to the measurement record for electron positions, and 
a Fourier transform yields the energy spectrum. As a concrete example 
we have illustrated the experimental setup for this approach using $^{40}{\rm 
Ca}^+$ ions, for which we find that the constraint set by trap decay time 
should permit electron propagation for as long as $\sim 0.1$ seconds to be 
studied.

When only the energy spectrum is required,  no additional complexity overhead is 
added in that we do not need to  simulate the dilatation operation  on 
the trapped ions explicitly (it is unitary and does not affect the spectrum). 
 If the wave function is also required,  only a simple unitary rotation on the 
initial state is necessary. Moreover, because of the rescaling of time the 
runtime of the simulator is much shorter than the timescale for evolution of the 
electron gas, minimizing the susceptibility of the quantum simulator to external 
noise and decoherence. 

Straightforward extension of this proposal can 
incorporate different geometries of the trapped ions; for example ion 
chains~\cite{Benassi2011}, ion rings~\cite{Wang2015}, and even periodic 
lattices~\cite{Lee2009,Tarruell2012} are feasible. 
Particularly interesting applications involve trapped ions loaded in 
two-dimensional honeycomb lattices.  This 
would permit emulation of electrons in graphene-like structures,
allowing massless Dirac quasiparticles and associated topological 
phases to be studied \cite{Polini2013}.

This technique can  be 
extended to many other interesting systems to solve the problem of scale 
difference between the kinetic and position-dependent terms or, more generally, 
the problem of terms having different powers of momentum and position 
dependence.

{\it Acknowledgments.}{\bf--}
This work is supported by the Basque Government (Grant No.~IT472-10), 
the Spanish MICINN (Project No.~FIS2012-36673-C03-03), and the Basque Country 
University UFI (Project No.~11/55-01-2013).  Partial support was provided by 
LightCone Interactive LLC.




\end{document}